\documentclass[twocolumn,showpacs,preprintnumbers,amsmath,superscriptaddress,amssymb,bbold]{revtex4-1}
\usepackage{appendix}
\usepackage{epsfig,amsopn}
\usepackage{graphicx}
\usepackage{epstopdf}
\usepackage{sidecap}
\usepackage{amsmath,amssymb}
\usepackage{amsthm}
\usepackage{enumerate}
\usepackage{bbold}
\newcommand\bea{\begin{eqnarray}}
\newcommand\eea{\end{eqnarray}}
\newcommand\beq{\begin{equation}}  
\newcommand\eeq{\end{equation}}

\begin{document}
\title{Fabry-Perot interferometry in Weyl semi-metals}
\author{Dibya Kanti Mukherjee}
\affiliation{Harish-Chandra Research Institute, Homi Bhabha National Institute, Chhatnag Road, Jhunsi, Allahabad 211 019, India}
\author{Sumathi Rao}
\affiliation{Harish-Chandra Research Institute, HBNI, Chhatnag Road, Jhunsi, Allahabad 211 019, India}
\author{Sourin Das}
\affiliation{Department of Physical Sciences, IISER Kolkata, Mohanpur, West Bengal 741246, India}
\affiliation{Department of Physics and Astrophysics, University of Delhi, Delhi 110007, India}
\begin{abstract}

We show that the electrical transport across a minimal model for a time-reversal symmetry(TRS) breaking Weyl semi-metal (WSM)  involving two Weyl nodes can be interpreted as an interferometer in momentum space. The interference phase depends on the distance between the Weyl nodes ($\vec{\delta k}$) and is
{\it anisotropic}. It is further shown that a minimal inversion symmetry broken model for a WSM with four Weyl nodes effectively mimics a situation corresponding to having two copies of the interferometer due to the presence of an orbital pseudo-spin domain wall in momentum space. We point out that the value of the $\delta k$ and consequently the interference phase can be tuned by driving the WSMs resulting in oscillations in the two terminal conductance measured in the direction of splitting of the Weyl nodes.

\end{abstract} 
\maketitle
\section{Introduction}

Although Weyl semi metals\cite{Murakami2007,Vishwanath2011,Burkov2011a, Burkov2011b,Zyuzin2012a,Hosur2012} have often been advertised as the three dimensional analogues 
of graphene\cite{graphene}, they are actually quite different.
 In the WSM,
the excitations at the nodes - the points where the conduction and valence band touch -  are chiral, which
means that their spins are fixed either along or opposite to their direction of motion, in contrast to graphene where
both spins are allowed.
The  chiral nodes also form a pathway through the bulk for unusual surface states which form 
disjointed Fermi arcs\cite{Vishwanath2011}. The direct evidence\cite{Xu2015a,Xu2015b,Lv2015a,Lv2015b,Lu2015,Jia2016} for these Fermi arcs have led to 
many more theoretical proposals\cite{Vazifeh2013,Khanna2014,Uchida2014,Burkov2015a,Burkov2015b,Goswami2015,Baum2015,Khanna2016,Behrends2016,Rao2016,Baireuther2016a,Tao2016,Chen2016,Marra2016,Li2016,Baireuther2016b,Madsen2016,Obrien2017,Khanna2017,Bovenzi2017,Mukherjee2017} in the last couple of years in this field.

The effect of periodic driving or irradiation of condensed matter systems giving rise
to new topological phases in Dirac materials, is another direction of research which has been pursued vigorously in the last few years. In particular, 
the physics of inter-nodal  Andreev reflection was studied recently\cite{Khanna2016} where, a geometry with a WSM sandwiched
between $s$-wave superconductors was studied.  It was shown  that the  Josephson current oscillates and undergoes zero-pi transitions  as a function
of the internodal distance which could be tuned by shining light on the
material\cite{Khanna2017}. More recently, this physics was also probed in a study of transport through a WSM quantum dot sandwiched between a normal metal and a superconductor and the interplay between Klein tunneling and the chirality induced effects were highlighted\cite{Mukherjee2017}.

\begin{figure}
\centering
\includegraphics[width=0.45\textwidth]{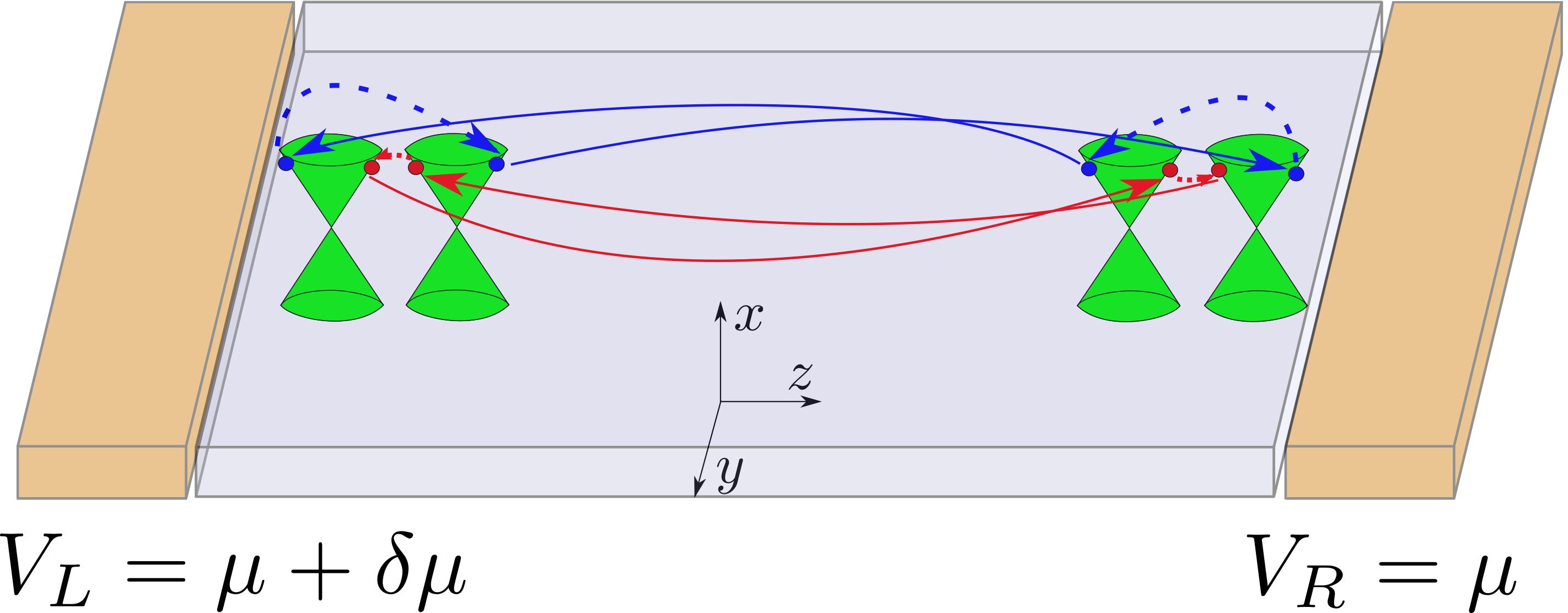}
\caption{ (Color online) Diagrammatic representation of the interference process. The dotted lines near the boundary of the system demonstrates the inter-nodal scattering wheras the continuous lines show that within the WSM bulk there is no further backscattering from one node to the other. The total phase picked up during one closed loop is given by $2k_0L$ as discussed in the main text.}
\label{figtrs}
\end{figure}

Unlike the previously studied  geometries involving superconductors, here we study the low energy transport  of a Weyl semimetal sandwiched between two {\it normal leads} to obtain a clear physical understanding of the mysterious looking oscillations in superconductor-normal hybrid junctions studied\cite{Mukherjee2017,Khanna2017,Khanna2016}. We note that the reflection that takes place  at the edges of the sample can be interpreted in terms of 
 a beam splitter
in momentum space which leads  to similar  oscillations in the normal current  as a function of $k_0 L$, where $2k_0$ is the distance between the two Weyl nodes in momentum space and $L$ is the length of the sample.  Here,  the spin texture and the chirality  of the excitations play a crucial role. As in the previous cases, the  oscillations are highly anisotropic
with the maximum amplitude seen in the current measured along the same direction as the separation of the Weyl nodes and vanishing in the direction perpendicular to the separation of the nodes. 


\begin{figure*}[ht]
\centering
\includegraphics[width=0.95\textwidth]{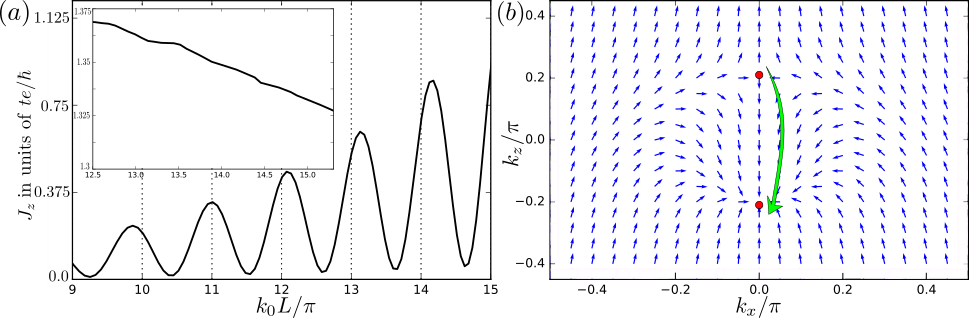}
\caption{ (Color online) (a) Oscillations in the current along the direction in which the Weyl nodes are split in the momentum space for the TR symmetry broken WSM bulk as a function of $k_0L/\pi$. The dotted lines  show the theoretically predicted periodicity. The parameters used are $t = 1$, $\lambda_{SO}=\lambda_z=0.5t$,$\epsilon=6t$,$\mu_L=0.3t$,$\mu_R=0$. The size of the WSM bulk, $L=62$, includes the lead sites at the two ends. The inset shows the behaviour of the current in the perpendicular direction with the same parameter values. It is quite clear that this does not show any oscillations as a function of $k_0L/\pi$. (b) Spin textures of the second band in the $k_x$-$k_z$ plane, with the Weyl nodes at $k=k_0\hat{z}$ and $k=-k_0\hat{z}$ indicated by red dots. Possible scattering process for a forward moving spin up electron is shown. The spin texture is symmetric under $\{k_x,\sigma_x\}\leftrightarrow \{k_y,\sigma_y\} $. The length of the unit cell is taken to be equal to 1.}
\label{figone}
\end{figure*}

There have been several  experimental and theoretical studies\cite{Hsieh2009,Chen2009,Zhu2011,Wray2011,Xu2012} of spin textures in topological insulators (TIs)\cite{Hasan2010,Qi2011} and in other Dirac materials\cite{Wehling2014}. 
However, much less attention has been paid to the study of the  emergent  pseudo-spin or orbital degree of freedom that appears in many of these materials. 
A notable exception are the works by Roy {\it et al}\cite{Roy2016a,Roy2016b},
which show that the orbital pseudo spin polarization\cite{FanZhang} of surface states of three dimensional topological insulators can have strong  physical manifestations.  Tunneling between two surfaces of topological insulators were show to be suppressed due to mismatch of  the orbital pseudo spin on the two surfaces even though all other dispersing degrees of freedom allow tunneling.
For a time-reversal symmetry breaking WSM whose minimal model has just two Weyl  nodes, the identification of the scale $\delta k$ with $2k_0$ is obvious, but for a time-reversal invariant, inversion symmetry broken WSM, the minimal model has
four Weyl nodes. Hence, naively the identification of $\delta k$ is not obvious. However, we shall see that
analogous to the behaviour in topological insulators,  orbital  pseudo spin conservation comes to the rescue. Unlike the spin whose polarization is tied to the direction of motion of the quasi-particle,  the polarization  of the orbital pseudo-spin turns out to be independent of the momentum in the inversion symmetry broken WSM.  
Hence, conservation of pseudo spin, allows for scattering only between a single pair of Weyl nodes, and 
we show that we get $\delta k L$ oscillations for an appropriate $\delta k$ even for inversion symmetry broken WSMs.

The paper is divided into several sections.  In section II, we introduce the TRS broken WSM model and we calculate the current as a function of the internodal distance in the NWN geometry for a fixed chemical potential difference across the sample. We demonstrate the periodic oscillations and provide the mechanism behind the same. In section III, we repeat the same process for an inversion symmetry broken WSM and introduce additional selection rules which dictate the scattering processes as TRS is preserved. Lastly, in section IV, we discuss the robustness of these oscillations and propose experimental setups that could probe into this kind of transport.

\begin{figure*}
  \centering
  \includegraphics[width=0.99\textwidth]{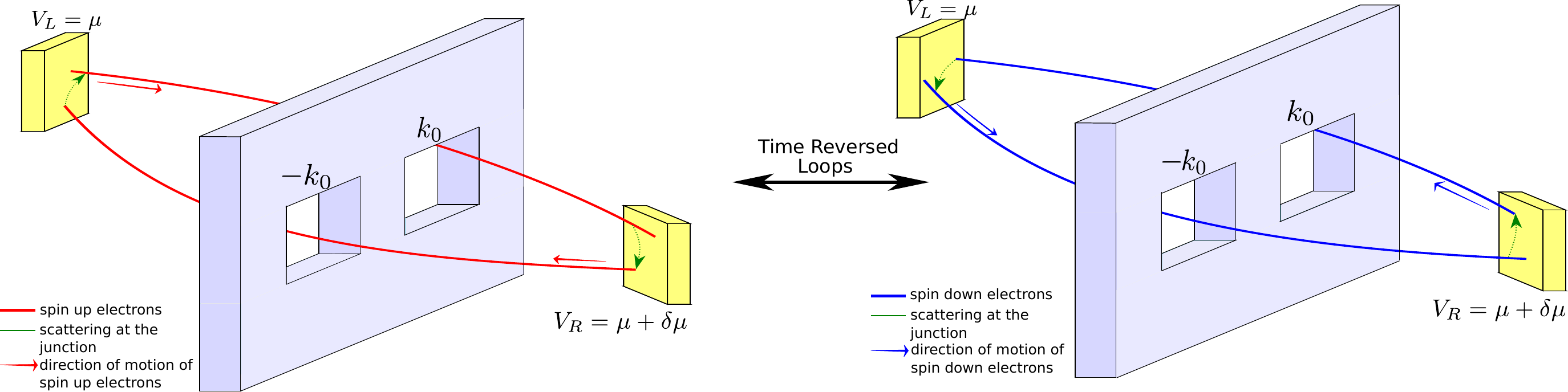}
  \caption{ (Color online) Diagrammatic representation of the interference process. The dotted lines near the boundary of the system demonstrates the inter-nodal scattering whereas the continuous lines show that within the WSM bulk there is no further backscattering from one node to the other. The red loop denotes the momentum loop for the spin up electrons and the blue
  loop denotes the same for the spin down electrons. The total phase picked up during one closed loop is given by $2k_0L$ as discussed in the main text.}
  \label{figthree}  
\end{figure*}

\section{Time-reversal breaking WSM}


In this section, we will study the current through a junction of a TRS breaking WSM of length $L$ between two normal leads.


The WSM is modeled by the standard Hamiltonian for a three dimensional topological insulator in the $Bi_2Se_3$ class with a time-reversal breaking perturbation($b_z$) added to make it a WSM
\cite{Vazifeh2013,Khanna2014} -
\begin{align}\label{eq:ham}
  H_{0} =& \epsilon_k \tau_x - \lambda_z \sin k_z \tau_y  \nonumber \\
  & -\lambda_{SO} \tau_z \left( \sigma_x\sin k_y  - \sigma_y \sin k_x\right) + b_z \sigma_z, 
\end{align}
where $\epsilon_k = \epsilon - 2t \sum_i \cos k_i$ is the kinetic energy,  $\tau$ represents the orbital (pseudo-spin) degree of freedom, $\sigma$ represents the spin and  $\lambda_{SO}$, $\lambda_z$ are  the strengths of the spin-orbit coupling. Here and throughout the rest of the paper, we set the lattice spacing($a$) to be unity. All momenta are expressed in units of $1/a$ and all lengths in unit of $a$.

This is a 4 band model where,
in the limit where $\lambda_z\ll \epsilon -6t \ll b_z$, 
the two middle bands touch at $(0,0,\pm k_0)$ forming a pair of Weyl nodes.  Here $k_0$  is defined via $tk_0^2 = b_z - \epsilon +6t$, and the top-most and lower most bands are far from the touching point and can be ignored at low energies. Thus the model reduces to the the two band model studied in detail in Ref.\onlinecite{Uchida2014}.

To compute the current, we will use the lattice version of this four band model. We choose the chemical potential $\mu_W$ (measured from the Weyl node) to be sufficiently small, so that the states belonging to the two Weyl nodes are well-separated.
Note that our choice of the Weyl nodes along the $k_z$ axis implies that the surface states for this model appear on  the surfaces perpendicular to the $x$ or  $y$ axes.
We will compute conductances through the WSM both along the $z$ direction, which is the direction of the separation of the Weyl nodes and along the $x$ direction which is perpendicular to the separation.  We will compute it explicitly using the lattice version of Eq.1 using the Green's function technique.
The Green's function for the isolated Weyl semimetal is constructed as $G_{W}(\omega) = (\omega - H_{W})^{-1}$. The two leads on either side of the WSM are coupled via self-energy terms in the full Green's function. Since the attached  leads are  expected  to have very large bandwidths compared to that of the WSM,
these self energies are taken to be constant matrices diagonal in the $\sigma$ and $\tau$ indices. The full Green function is  then given by
\begin{equation}
  \begin{aligned}
    G(\omega) = (G_W^{-1}-\Sigma_L-\Sigma_R)^{-1}~.
  \end{aligned}
\end{equation}
Here $\Sigma_j = -i\tilde{t}_j^2\pi\delta_{\sigma,\sigma'}\delta_{\tau,\tau'}$ with $\tilde{t}_j$ being the hopping amplitude from the Weyl semimetal to the $j^{th}$ lead. The current is then obtained as 
\begin{equation}
  \begin{aligned}
    \langle J\rangle = \frac{2\pi e}{\hbar^2}\int d\omega Tr[G^{\dagger}(\omega)\Gamma_RG(\omega)\Gamma_L](f_R(\omega) - f_L(\omega))~.
  \end{aligned}
\end{equation}
Here $\Gamma_j = i(\Sigma_j - \Sigma^{\dagger}_j)$ and $f_j$ is the Fermi function of the $j^{th}$ lead. We assume that biasing the system does not change the band structure of the WSM bulk. This assumption should be valid in the linear response regime. Hence, throughout this paper, we only consider cases where the biasing is small compared to the bandwidth of the system.

In Fig.\ref{figone}(a), we show that the current with fixed bias, when measured along the $z$ direction shows oscillations as a function of $k_0L$, whereas no oscillations are seen in the $x$ direction as shown in the inset of Fig.\ref{figone}(a). It is to be noted here that in the transverse direction, most of the contribution to the current comes from oblique incidence since the dispersion near $k_y=k_z=0$ is gapped. The assumption here is that the leads are sufficiently large in the $y$ and $z$ directions.
In our computation, we take the two lateral directions to be periodic and we sum over 200 transverse modes in each periodic direction.

In order to explain the oscillations, we first compute the spin texture of the low energy bands near the two nodes. Near the Weyl nodes, the low energy two component Hamiltonian can be written as
\begin{align}\label{eq:HWSM}
H_{\text{WSM}} = \epsilon_k{\tilde \sigma}_z -\mu_W + \lambda (k_x{\tilde \sigma_x} + k_y{\tilde \sigma_y})
\end{align}
after a unitary transformation and projection onto the low energy subspace(see Appendix A) of the Hamiltonian in Eq.\ref{eq:ham}\cite{Khanna2016}.
Here  $\epsilon_k = (\hbar^2/2m_W) (k_x^2+k_y^2+k_z^2-k_0^2)$ is  the kinetic energy, $\mu_W$ is the chemical potential measured from the Weyl node and $m_W$ is the effective mass and we have chosen the effective spin $\tilde\sigma = -\sigma$, in terms of the spin of the original Hamiltonian. Note that the orbital pseudo spin is no longer a good quantum number in this effective model. 
Also note that this model, when projected to the low energy subspace, breaks inversion symmetry\cite{Bovenzi2017}. In the lattice model, we have computed the  spin textures of the lower (filled) band of the  two middle bands of the 4 band model and as can be seen in Fig.\ref{figone}(b), the spin texture  at $k=k_0$ is not opposite to the spin texture at $k=-k_0$. 
Thus, in terms of the two dimensional low energy reduced subspace, the Hamiltonian for the model is given by
\beq
H = v_xk_x{\tilde\sigma}_x +v_yk_y{\tilde\sigma}_y+ sv_zk_z{\tilde\sigma}_z
\eeq
where $v_x=\lambda,v_y=\lambda$ and $v_z=\hbar^2k_0/m_W$, ${\tilde\sigma}_i$ are the effective two component spins and $s=\pm 1$ for the two nodes. 
We also note that although the spin textures are not opposite, the topological quantum numbers as computed either as the sign of the product of the three velocities ($sv_xv_yv_z$) or from  the integrated Berry curvature ($2\pi s$) around the nodes is  clearly opposite for the two nodes.

We can now easily understand the oscillations in $k_0L$  by using the scattering matrix formalism in the reduced two band model(see Appendix B). We  restrict ourselves to small $\mu_W$, so that the two Weyl nodes are disconnected at the Fermi energy and obtain the wave-functions by using the scattering matrix formalism.
The important point to note is that  since in a two band model, spin conservation during the scattering processes at the leads implies that even normal reflection at any interface is inter-nodal\cite{Uchida2014,Khanna2016}, multiple reflections between the two leads can give rise to $k_0L$ oscillations.

As can be seen from the Fig. \ref{figthree}, an electron with spin up entering the WSM from the left lead with an energy close to the Fermi energy of the WSM bulk is automatically forced to occupy a forward moving quantum state close to the Weyl node at $\vec{k} = (0,0,k_0)$ due to the spin texture of the low lying bands. As the electron travels through the bulk, its chirality is preserved. However, at the right junction, it can either leak into the right lead or can backscatter into a quantum state near the other Weyl node at $\vec{k} = (0,0,-k_0)$ moving to the left. The electron then travels to the left junction and gets reflected back to the original Weyl node, thus performing a closed loop evolution thereby introducing an interferometric phase. It is to be noted that the closed loop actually comprises of evolution both in real space (traversal from one lead to the other with a fixed momentum) and momentum space (scattering from one Weyl node to the other) processes. In higher orders of reflection amplitude, this fundamental loop repeats itself. The final result is a geometric progression in this loop which can be summed over. It can be easily shown that the phase picked up by the electron during one such loop is $= 2k_0L + \delta\mu L/v_z$ where $\delta\mu$ is the chemical potential of the bulk WSM. This phase relation becomes equal to $2k_0L$ as the chemical potential is tuned closer to the Fermi energy of the WSM bulk. The same structure for the amplitudes is repeated by an electron with down spin but in a time reversed fashion as depicted in Fig. \ref{figthree}. 

It is to be noted here that the interference happens between the dominant unscattered classical path from the left lead to the right and the subleading loops. This can be thought of as a Fabry-P\'erot interferometer with one arm being weakly coupled, where the coupling strength is equal to reflection strength. Since the final states of the spin up and spin down sectors are orthogonal to each other, we do not get any interference from these two time reversed partners.

For a standard metal sandwiched between two leads the Fabry-Perot oscillations have a period given by superposition of $2k_F$ oscillations induced by each Fermi surface, in a multiband picture, which appear at the Fermi level. Hence the oscillation period is connected to backscattering events within each Fermi surface. Such Fabry-Perot resonances have been studied in electron tunneling through double barriers on
the surface of a topological insulator\cite{wu2010} and even experimentally seen  in trilayer graphene\cite{campos2012} and in suspended single layer graphene\cite{oksanen2014}.  In sharp contrast,  in the case of a Weyl semimetal, the oscillation period is dictated  primarily by internodal momentum differences and not the intranodal $2k_F$ which is very small at low doping. This is because, as indicated earlier, intranodal backscattering is forbidden by spin selection rules. This fact is independent of the details of the band structure of the Weyl semimetal.

\begin{figure}[t]
\centering
\includegraphics[width=0.49\textwidth]{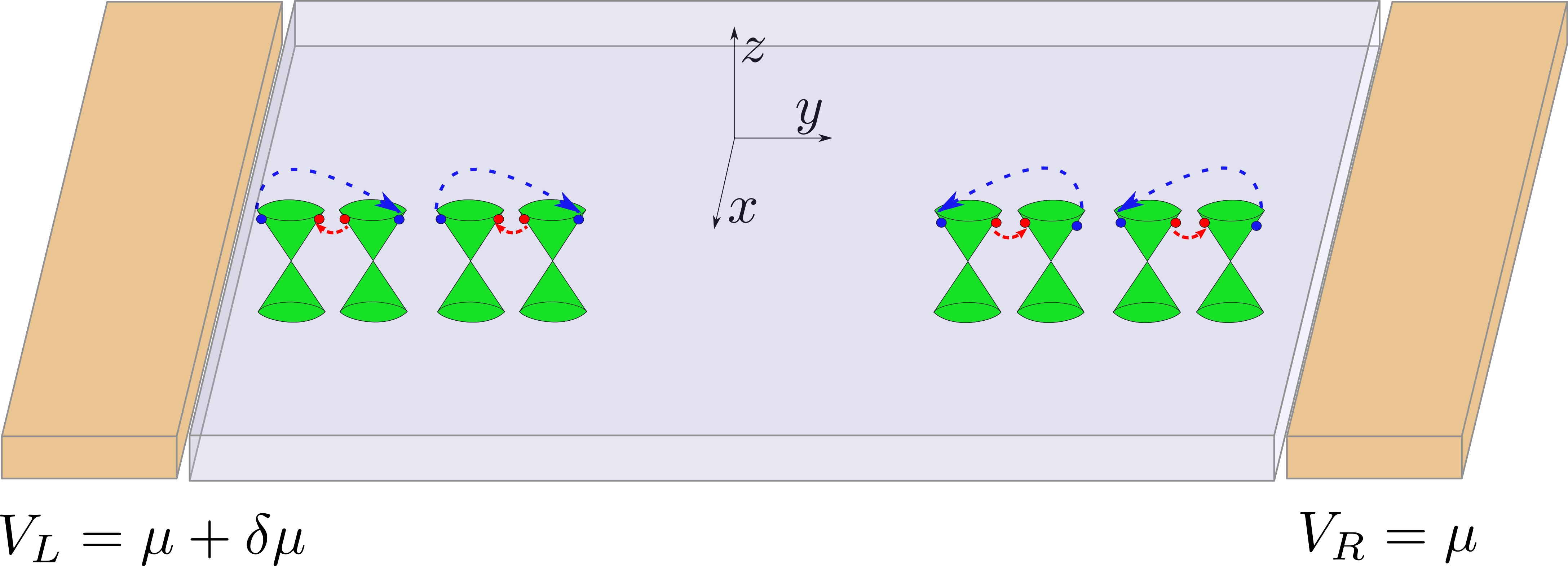}
\caption{ (Color online) Diagrammatic representation of the interference process for the inversion symmetry broken WSM bulk. The dotted lines near the boundary of the system demonstrates the selective inter-nodal scattering, where spin and orbital degrees of freedom are preserved(see main text for details). We have not explicitly shown the traversal of the electrons within the semi-metal, since it would involve twice as many lines as shown in Fig. 1. As  noted in the main text, this setup essentially reduces to two copies of the TRS broken WSM case due to preservation of the pseudospin degree of freedom.}
\label{figinv}
\end{figure}

\section{Time-reversal invariant, inversion symmetry broken WSM}

In this section, we study the same geometry as before, but with the time-reversal symmetry broken WSM
replaced by a time-reversal invariant, inversion symmetry broken WSM. A minimal model
in this class\cite{Chen2016,Khanna2017} is given by
\begin{equation}\label{eqinv}
  \begin{aligned}
    H_{inv} = \lambda\sum_{\alpha=x,y,z}\sigma_{\alpha}\textrm{sin}k_{\alpha} + \sigma_y\tau_yM_{\textbf{k}} - \mu 
  \end{aligned}
\end{equation}
where $M_{\textbf{k}} = (m+2-\textrm{cos}k_x-\textrm{cos}k_z)$. This model describes a trivial insulator when $m>\lambda$. The bulk gap closes at $m=\lambda$ and two Dirac nodes appear at $\textbf{k}=(0,\pm\pi/2,0)$. For $m<\lambda$, each of the Dirac nodes split into two Weyl nodes along the $k_y$ axis, so that the model has four Weyl nodes along the $k_y$ axis at $k_y = \pm (\pi - k_0)$ and $k_y = \pm k_0$ where $k_0 = \sin^{-1}(m/\lambda)$. 

\begin{figure*}[t]
\centering
\includegraphics[width=1.0\textwidth]{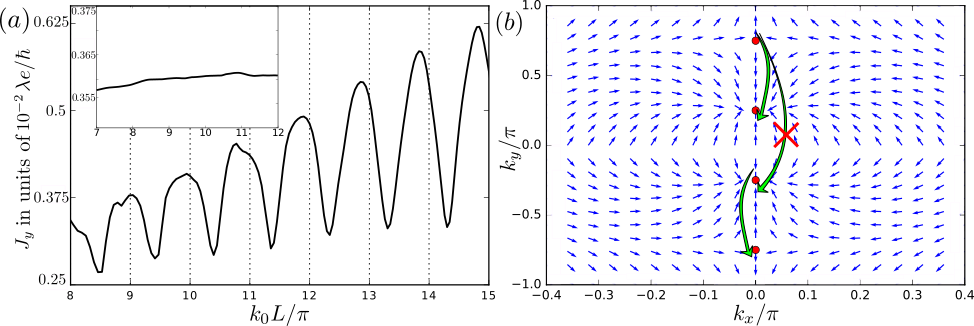}
\caption{ (Color online) (a) Current along the direction in which the Weyl nodes are split in the WSM bulk for the inversion symmetry broken WSM as a function of $k_0L/\pi$. The dotted lines clearly indicate the predicted periodicity due to internodal scattering. The periodicity is a function of only one of the relevant momentum scales of the problem(see main text). The parameters used are $\lambda = 1$, $\mu_L = 0.1\lambda$, $\mu_R = 0$. The number of sites is kept fixed at $L=62$ including the two lead sites. The inset shows the variation of current in the perpendicular direction for the same set of parameter values. As expected, it does not exhibit any periodicity. (b) Spin textures of the second band with the Weyl nodes indicated by red dots for the inversion symmetry broken WSM. Possible scattering process for a forward moving spin up electron is shown. The red cross indicates the absence of the corresponding process due to preservation of the orbital degree of freedom. The spin texture is symmetric under $\{k_x,\sigma_x\}\rightarrow \{k_z,\sigma_z\} $.The effect of a parameter scrambling the pseudospin has been studied in Sec. IV.}
\label{figtwo}
\end{figure*}

We  now compute the current when the leads are placed at $y=-L/2$ and $y=L/2$ - i.e., along
the direction of the Weyl nodes in momentum space, and also when the leads are along the 
$x$ axis, perpendicular to the Weyl nodes. The lead biases are kept fixed at $\mu_L=0.1\lambda$ and $\mu_R=0$. The results for the y and x axes are shown in Figs \ref{figtwo}(a) and its inset. 
Surprisingly, we find that our results are similar to those found in the time-reversal breaking 2 Weyl node WSM, although here there are four Weyl nodes, and naively, the reflection at each of the junctions with the normal leads can lead to reflection from 2 other possible Weyl nodes, if only spin conservation is taken into account, as can be seen from the spin structures plotted in Fig.\ref{figtwo}(b). However, when we taken into account the conservation of pseudo spin (also called orbital spin), this is no longer true. 

\begin{figure}[b]
\centering
\includegraphics[width=0.45\textwidth]{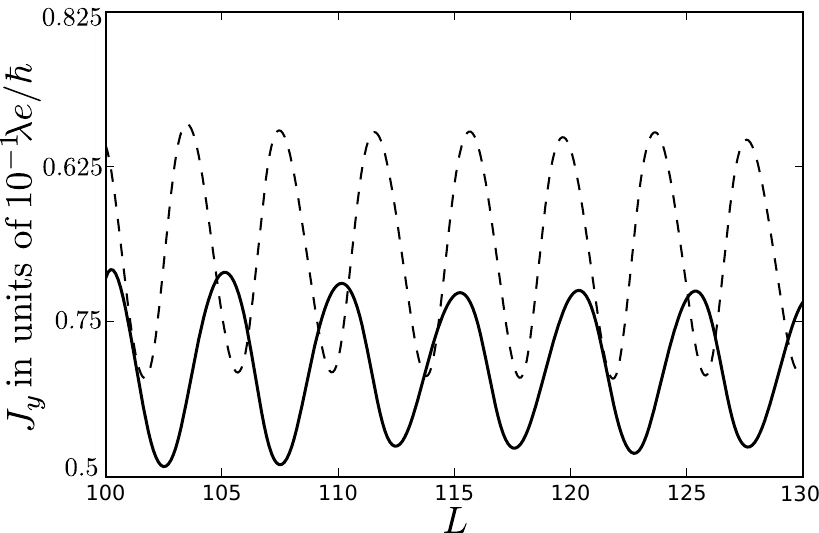}
\caption{ (Color online) Variation of current as a function of the number of lattice sites. The dashed line has a periodicity of 4 sites whereas the continuous line has a periodicity of 5 sites. The parameters used are $\lambda = 1$, $\mu_L = 0.1\lambda$, $\mu_R = 0$.}
\label{figlength}
\end{figure}

Just focusing on the $k_y$ axis, the Hamiltonian in eq.(\ref{eqinv}) reduces to 
\beq
h(k_y)=(\lambda \textrm{sin}k_y + m\tau^y)\sigma^y=\pm(\lambda \textrm{sin}k_y+m\tau^y).
\eeq
We note that the lower conduction band has lower value of energy ($\lambda\textrm{sin}k_y-m$) and is thus energetically constrained to have $\langle \tau^y\rangle = -1$ ($+1$) when $k_y>0$ ($k_y<0$). This is in fact true for any other axis parallel to the $k_y$ axis. The pseudospin $\langle\tau_y\rangle$ thus forms a domain wall in momentum space. This argument can be repeated in its entirety for the higher among the two valence bands and we  find that $\langle\tau^y\rangle$ is constrained to be equal to $-1$ when $k_y>0$ and $\langle\tau^y\rangle = +1$ when $k_y<0$. This is exactly the same as the values for the lower conduction band. This prevents $\langle\tau^y\rangle$ from changing even for the two middle bands across the band touching points. 

This implies that for the middle two bands in this four band model, only pseudospin preserving scattering processes can occur between the Weyl nodes at $k_y =\pi - k_0$ and $k_y=k_0$ or between the nodes at
$k_y = - \pi + k_0$ and $k_y = -k_0$.  This explains why there is a single relevant scale $k_0$ in the scattering
at the junctions even though there are multiple momentum scales in the problem (all possible differences between the nodes, or at least two possible momenta, even if we impose spin conservation).

This is further demonstrated in Fig. \ref{figlength}. Here, the value of current along the direction of separation of Weyl nodes is plotted as a function of the length of the system by fixing $k_0$. Let us consider an electron with $\langle\sigma_y\rangle = 1$ moving along the $k_y$ axis with momentum $k_y=\pi-k_0+\delta$ where $\delta \ll k_0$. If spin is conserved, this electron can backscatter to two possible quantum states - one at $k_y=k_0-\delta$ and the other at $k_y=-k_0-\delta$ at a junction. The separation in the BZ of the first state with the incoming electron is equal to $\pi - 2k_0$ wheras that of the second state is equal to $\pi$.

In Fig. \ref{figlength} the phase picked up for the chosen set of parameter values is equal to $(\pi - 2k_0)L \approx 0.5\pi L$ and $\pi L$ in the other. Hence, for the first process, we expect a periodicity of $\approx$ 4 sites whereas, and for the second case, we expect the periodicity to be 2 as a function of the length($L$). As we vary the length of the system, we find that there is a repeated pattern after the length of the system is varied by 4 sites. This is given by the dashed lines in Fig. \ref{figlength}. The entire process is repeated by changing the parameters of the Hamiltonian such that $(\pi - 2k_0)L \approx 0.4\pi L$. This, as expected, produces a periodicity of $\approx$ 5 sites as is seen from the continuous line in Fig \ref{figlength}. This conclusively establishes that backscattering is allowed only if pseudospin($\tau$) is preserved. Note also that  unlike in Figs. (3) and (5),
where the amplitude of the oscillations  increases as a function of $k_0L$, here the amplitude of the oscillation has saturated since
we have used a larger length $L$.

We also note that this explains the earlier results on the oscillations in the Josephson current in the time-reversal breaking inversion symmetry conserved WSM\cite{Khanna2017} where the oscillations only depended on a single relevant scale $k_0$ identified as the distance between the nodes either both on the left or both on the right of the $k_y=0$ axis. A similar selection rule was also noted in the appendix of Ref.\cite{Mukherjee2017}. However, the role of the pseudospin conservation was not noted.

Thus, although in principle, the  description of the multiple scattering processes in the TR-symmetric, inversion symmetry broken WSM could have been quite different from that of the TR broken WSM due to the presence of additional momentum scales, the selection rule introduced by the preservation of the pseudo-spin degree of freedom($\tau$) during scattering processes reduces this setup to be exactly two copies of the previously discussed interferometer.

\begin{figure}[t]
\centering
\includegraphics[angle = 0, width=0.45\textwidth]{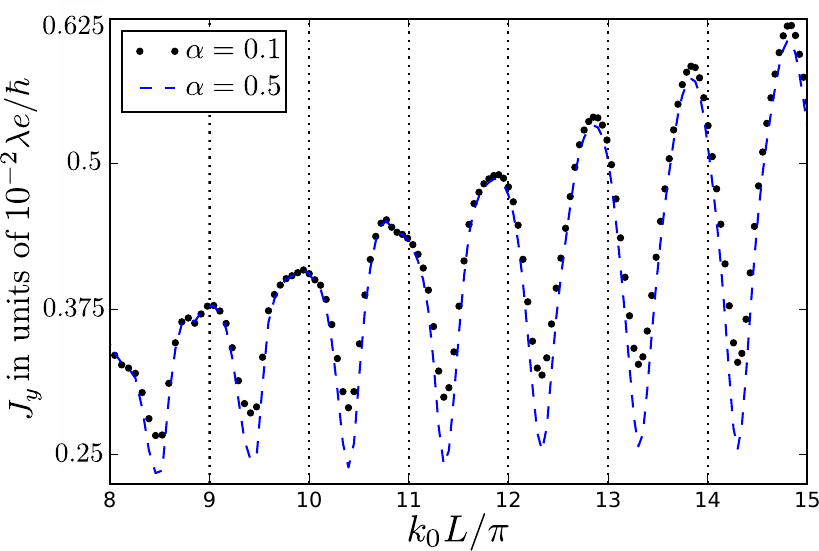}
\caption{ (Color online) Variation of current for different values of the scrambling parameter $\alpha$.}
\label{figscramble}
\end{figure}

\section{Discussion}

In the numerical results presented in the article, the self-energy terms considered are diagonal in the $\sigma$ and $\tau$ basis. Although it may be argued that both the conservation of pseudospin and its ${\bf k}$-independence are
model dependent statements, note that it is not unrealistic to expect that a WSM made by stacked layers of 2D topological
insulators would have a $k$-independent degree of freedom, analogous to the frozen $SU(2)$ degree of freedom established\cite{zhang2013} in three dimensional topological insulators. Also, to check for the model independence of our results,
we study the possible effects of off-diagonal self-energy by introducing a minimal perturbation of the diagonal $\Sigma$ given by $\Sigma = -i\tilde{t}_j^2\pi\delta_{\sigma,\sigma'}\otimes(\delta_{\tau,\tau'}+\alpha\tau_x)$. Here, $\alpha$ is the parameter that scrambles the pseudospin($\tau$) inside the leads. As can be seen from fig.(\ref{figscramble}), we find that the period of oscillations does not depend on $\alpha$, though the amplitude has some dependence. In fact the same analysis for the TR symmetry broken model exhibit no dependence of the current on the $\alpha$ parameter. Hence, it has not been shown here. We thus conclude that the oscillations are quite robust and do not exactly depend on how precisely the WSM bulk is connected to the leads as well as on the nature of the leads and are a consequence of the chirality of the bulk Weyl nodes. 

As far as the relevance of our findings to experimental observations are concerned, it should be noted that there are various theoretical proposals for manipulation of the bulk band structure in a WSM. For us, the parameter of relevance is the separation of the Weyl nodes and even for non-collinear modes, or for models with many more Weyl nodes, all that would matter would be the projection of the separation along the
direction of the leads. One way of changing the separation of the Weyl nodes is by shining high frequency light which has been explored in Ref.\cite{Khanna2017}. Another interesting way of doing the same could be by straining the sample\cite{Franz2016}.  In fact, an important point to note is that our model requires creating junctions of WSM with normal leads which should be much simpler experimentally  than creating junctions of WSM with superconductors. The direction of separation of the Weyl nodes can be obtained from analising ARPES data of the surface Fermi arc states. Note also that unlike the $K-K'$ scale in graphene, the $k_0$ scale in Weyl semimetals is much smaller and does not require
very large momentum transfer. Hence, small wiggles in the interfaces will not be able to wash out
the signal. Moreover, even atomically sharp junctions are not a bottle-neck with the present day technology  of growing thin films of topologically materials\cite{bansal2011,kim2017}. 

To summarise, we have proposed a setup to explore the nature of chiral nodes in WSM by studying transport through a normal-metal-WSM-normal-metal heterostructure. We have considered the minimal models of both inversion symmetry broken and TR symmetry broken WSM-s. We have identified the oscillations that we observe in each of the cases to be closely related to the preservation of the symmetries  of the microscopic models. For the TR broken WSM, spin plays the role of the conserved quantity that leads to selection rules for scattering processes in the boundary. For the inversion symmetry broken WSM, the orbital degree of freedom plays an important role along with the spin. In fact, these selection rules lead to the interferometry that we have explored in the previous sections.  

\begin{acknowledgments}
We would like to thank G. Murthy for useful discussions and comments.
The research of DKM was supported in part by the INFOSYS scholarship for senior students.  
\end{acknowledgments}


\appendix

\section{Identification of low-energy subspace}

In this appendix we elaborate on the methodology for identification of the low energy subspace for the TR symmetry broken WSM, with the Hamiltonian 
\begin{align}
  H_{0} =& \epsilon_k \tau_x - \lambda_z \sin k_z \tau_y  \nonumber \\
  & -\lambda \tau_z \left( \sigma_x\sin k_y  - \sigma_y \sin k_x\right) + b_z \sigma_z .\nonumber
\end{align}
We divide the Hamiltonian into three parts:
\begin{align}
  h_{1} =& \epsilon_k \tau_x - \lambda_z \sin k_z \tau_y  \nonumber \\
  h_{2} =& -\lambda \tau_z \left( \sigma_x\sin k_y  - \sigma_y \sin k_x\right) \nonumber\\ {\rm and}\quad
  h_{3} =& b_z \sigma_z~. \nonumber
\end{align}
We diagonalise $h_1$ using  the unitary operator:
\begin{align}
  U = \frac{1}{\sqrt{2}}
   \left( \begin{array}{cc}
            \mathbb{1} & \mathbb{1} \\
    e^{i\phi}\mathbb{1} & -e^{i\phi}\mathbb{1}
          \end{array} \right)~,
\end{align}
where $\phi = -{\rm tan}^{-1}(\lambda_z{\rm sin}k_z/\epsilon_k )$. Here, the unit matrix $\mathbb{1}$ is in the spin space. Hence,
\begin{align}
  U^{\dagger} h_1U=r(\textbf{k})\tau_z\mathbb{1}_{\sigma}
\end{align}
where, $r(\textbf{k})=(\epsilon_k^2+(\lambda_z{\rm sin}k_z)^2)^{1/2}$.
The second part of the Hamiltonian $h_2$, under this unitary operation, takes the form:
\begin{align}
  U^{\dagger} h_2U=-\lambda \tau_x \left( \sigma_x\sin k_y  - \sigma_y \sin k_x\right) \nonumber\\
\end{align}
Lastly, the third term of the Hamiltonian does not change because it involves only $\sigma$ matrices.

Hence, the full Hamiltonian, under the action of $U$ becomes:
\begin{align}
  H_U& = U^\dagger H_0U \nonumber \\
  =& \left( \begin{array}{cc}
            r(\textbf{k})\mathbb{1}+b_z\sigma_z & -\lambda(\sigma_x\sin k_y-\sigma_y\sin k_x) \\
            -\lambda(\sigma_x\sin k_y-\sigma_y\sin k_x) & -r(\textbf{k})\mathbb{1}+b_z\sigma_z \nonumber
          \end{array} \right).
\end{align}
When $k_x=k_y=0$, $H_U$ becomes quite trivial and it is easy to identify the low energy subspace to be the one with a relative sign difference between $r$ and $b_z$, ie, the middle $2\times 2$ block. This gives us the low-energy Hamiltonian:
\begin{align}
  H_L = \left( \begin{array}{cc}
             r(\textbf{k})-b_z & -\lambda(\sin k_y-i\sin k_x) \\
            -\lambda(\sin k_y+i\sin k_x) & -r(\textbf{k})+b_z \nonumber
          \end{array} \right)~.
\end{align}
This Hamiltonian will have Weyl nodes at $\textbf{k}_0=(0,0,\pm k_0)$ where $k_0 = \cos^{-1}(\frac{4t^2-\sqrt{Xb_z^2+\lambda_z^4}}{X})$ with $X = 4t^2-\lambda_z^2$.

%
%
%
%
%

\vspace{0.5cm}

\section{Scattering matrix calculation}

Let us try to demonstrate the spin selective reflection processes at the leads by a simple scattering matrix calculation. For simplicity, we have considered the transverse momentum to be zero. The Hamiltonian of the low energy two band effective model of the TR broken Weyl semimetal with two Weyl nodes at $k_z=\pm k_0$ is given by:
\begin{equation} H = 
  \left(
    \begin{array}{cc}
    k_z^2 - k_o^2 & 0 \\
     0 & - (k_z^2- k_o^2)
  \end{array} \right). \nonumber
\end{equation}
The eigenstates of this Hamiltonian are:
\begin{equation}\label{Weylstates}
  \left(
    \begin{array}{c}
      1 \\ 0
    \end{array}
  \right)e^{ip^+z},
  \left(
    \begin{array}{c}
      0 \\ 1
    \end{array}
  \right)e^{-ip^-z},
  \left(
    \begin{array}{c}
      1 \\ 0
    \end{array}
  \right)e^{-ip^+z},
  \left(
    \begin{array}{c}
      0 \\ 1
    \end{array}
  \right)e^{ip^-z}
\end{equation}
Here, $p^{\pm}=\sqrt{k_0^2\pm E}$ where E is the energy. $p^{\pm}$ are the momenta at the Weyl node at $k_z=k_0$ while the momenta $-p^{\pm}$ are located at the other node.
Let us consider another Hamiltonian describing a normal metal:
\begin{equation} H =
  \left(
    \begin{array}{cc}
    k_z^2 - \mu & 0 \\
     0 & k_z^2- \mu
    \end{array} \right). \nonumber
\end{equation}
The eigenstates of this Hamiltonian are:
\begin{equation}\label{normalstates}
  \left(
    \begin{array}{c}
      1 \\ 0
    \end{array}
  \right)e^{ipz},
  \left(
    \begin{array}{c}
      0 \\ 1
    \end{array}
  \right)e^{ipz},
  \left(
    \begin{array}{c}
      1 \\ 0
    \end{array}
  \right)e^{-ipz},
  \left(
    \begin{array}{c}
      0 \\ 1
    \end{array}
  \right)e^{-ipz}
\end{equation}
Here, $p=\sqrt{E+\mu}$ where E is the energy. The first two eigenstates in both Eqs.(\ref{Weylstates}) and (\ref{normalstates}) describe right moving states while the latter two describe left moving states.

Let us consider a process where an electron with momentum $=p^+$ is incident on a WSM-normal interface from the left. In such a process for $z<0$:
\begin{equation}
  \psi_W(z) = 
  \left(
    \begin{array}{c}
      1 \\ 0
    \end{array}
  \right)e^{ip^+z} + A
  \left(
    \begin{array}{c}
      1 \\ 0
    \end{array}
  \right)e^{-ip^+z}+B
  \left(
    \begin{array}{c}
      0 \\ 1
    \end{array}
  \right)e^{ip^-z}
\end{equation}
where $A$ and $B$ are respectively the amplitudes of internodal and intranodal reflections from the interface.
For $z>0$,
\begin{equation}
  \psi_S(z) = 
   C
  \left(
    \begin{array}{c}
      1 \\ 0
    \end{array}
  \right)e^{ipz}+D
  \left(
    \begin{array}{c}
      0 \\ 1
    \end{array}
  \right)e^{ipz}.
\end{equation}
We match the wavefunctions and their derivatives by the following equations:
\begin{eqnarray}
  \psi_W(z)|_{z=0} &=&\psi_S(z)|_{z=0} \nonumber \\
{\rm and} \quad 
  \sigma_z\partial\psi_W(z)|_{z=0} &=& \partial\psi_S(z)|_{z=0}.
\end{eqnarray}
One can solve these equations to find $B=D=0$ and $A=\frac{p^+-p}{p^++p}$,$C=\frac{2p^+}{p^++p}$. This can be repeated for an incident electron with momentum $=-p^-$ and the conclusion is the same. This simple exercise clearly demonstrates that the probablility of scattering to the same node is suppressed when compared to that to the opposite node. This results in the reflection amplitude picking up a phase equal to $(p^+-p^-)L\approx 2k_0L$ for small $E$ when this exercise is repeated for the other boundary of the NWN geometry at $x=L$. A similar analysis has been done in \cite{Khanna2016}.


\begin{thebibliography}{99}


\bibitem{Murakami2007} S. Murakami, New J. Phys. {\bf 9}, 356 (2007).

\bibitem{Vishwanath2011} X.~Wan, A.~M.~Turner, A.~Vishwanath and S.~Y.~Savrasov,
Phys. Rev. B {\bf 83}, 205101 (2011).

\bibitem{Burkov2011a}
A.~A.~Burkov and L.~Balents,
\newblock \prl\ {\bf 107}, 127205 (2011).

\bibitem{Burkov2011b}
A.~A.~Burkov, M.~D.~Hook and L.~Balents,
\newblock \prb\ {\bf 84}, 235126 (2011).

\bibitem{Zyuzin2012a}
A.~A.~Zyuzin, S.~Wu and A.~A.~Burkov,
\newblock \prb\ {\bf 85}, 165110 (2012).

\bibitem{Hosur2012}
P.~Hosur, S.~A.~Parameswaran and A. Vishwanath, \prl\ {\bf 108}, 046602 (2012).

\bibitem{graphene} For a review, see A. H. Castro Neto, F. Guinea, N. M. R. Peres, K. S. Novoselov and A. K. Geim, Rev. Mod. Phys. {\bf 81}, 109 (209).

\bibitem{chiral} H. B. Nielsen and M. Ninomiya, Phys. Lett. {\bf B}, 291(1980).

\bibitem{Xu2015a}
S.-Y.~Xu, I.~Belopolski, N.~Alidoust, M.~Neupane, G.~Bian, C.~Zhang, R.~Sankar, G.~Chang, Z.~Yuan, C.-C.~Lee, S.-M.~Huang, H.~Zheng, J.~Ma, D.~S.~Sanchez, B.~Wang, A.~Bansil, F.~Chou, P.~P.~Shibayev, H.~Lin, S.~Jia, and M.~Z.~Hasan, Science {\bf 349}, 613 (2015).


\bibitem{Xu2015b}
S.-Y.~Xu, N.~Alidoust, I. ~Belopolski, Z. ~Yuan, G.~Bian, T.-R. ~Chang, H. Zheng, V. N. Strocov, D.~S.~Sanchez, G.~Chang, C. Zhang, D. Mou, Y. Wu, L. Huang, C.-C.~Lee, S.-M.~Huang, B.~Wang, A.~Bansil, H.-T. Jeng, T. Neupert, A. Kaminski, H. Lin, S. Jia and M.~Z.~Hasan, Nat. Phys. {\bf 11}, 748 (2015).


\bibitem{Lv2015a}
B.~Q.~Lv, H.~M.~Weng, B.~B.~Fu, X.~P.~Wang, H.~Miao, J.~Ma, P.~Richard, X.~C.~Huang, L.~X.~Zhao, G.~F.~Chen, Z.~Fang, X.~Dai, T.~Qian and H.~Ding,
\newblock Phys. Rev. X {\bf 5}, 031013 (2015).

\bibitem{Lv2015b}
B.~Q.~Lv, N. ~Xu, H.~M.~Weng, J. Z. Ma,  P.~Richard, X.~C.~Huang, L.~X.~Zhao, G.~F.~Chen, C. E. Matt, F. Bisti, V. N. Strocov, J. Mesot, Z.~Fang, X.~Dai, T.~Qian, M. Shi and H.~Ding,
Nat. Phys. {\bf 11}, 724 (2015).


\bibitem{Lu2015}
L. ~Lu, Z. ~Wang, D. ~Ye, L. ~Ran, L. ~Fu, J. ~D. ~Joannopoulos and M.~ Soljacic, Science {\bf 349},  622 (2015).

\bibitem{Jia2016}
S. Jia, S.-Y. Xu and M. Z. Hasan, Nat. Mat. {\bf 15},  1140 (2016).

\bibitem{Vazifeh2013}
M.~M.~Vazifeh and M.~Franz,
\newblock \prl\ {\bf 111}, 027201 (2013).

\bibitem{Khanna2014}
U.~Khanna, A.~Kundu, S.~Pradhan and S.~Rao, \prb\ {\bf 90}, 195430 (2014).

\bibitem{Uchida2014}
S.~Uchida, T.~Habe, and Y.~Asano,
\newblock J. Phys. Soc. Jpn. {\bf 83}, 064711 (2014).


\bibitem{Burkov2015a}
A.~A.~Burkov,
Journal of Physics: Condensed Matter {\bf 27}, 113201 (2015).

\bibitem{Burkov2015b}
A.~A.~Burkov,
\prb\ {\bf 91}, 245157 (2015).


\bibitem{Goswami2015}
P.~Goswami, J.~H.~Pixley  and S.~Das Sarma, \prb\ {\bf 92}, 075205 (2015).

\bibitem{Baum2015}
Y.~Baum, E.~Berg, S.~A.~Parameswaran and A.~Stern, Phys. Rev. X {\bf 5}, 041046 (2015).

\bibitem{Khanna2016}
U. Khanna, D. K. Mukherjee, A. Kundu and S. Rao, \prb\ {\bf 93}, 121409(R) (2016).

\bibitem{Behrends2016}
J. Behrends, A. G. Grushin, T. Ojanen and J. H. Bardarson, \prb\ {\bf 93}, 075114 (2016).

\bibitem{Rao2016}
S. ~Rao, arXiv:1603.02821,  Jnl of Indian Institute of Science, 96(2), 145 (2016).

\bibitem{Baireuther2016a}
P. Baireuther, J. A. Hutasoit, J. Tworzydlo and C. W. J. Beenakker, New J. Phys. {\bf 18}, 045009 (2016).



\bibitem{Tao2016}
T. Zhou, Y. Gao and Z. D. Wang, \prb\ {\bf 93}, 094517 (2016).

\bibitem{Chen2016}
A. Chen and M. Franz, Phys. Rev. B {\bf 93}, 201105(R) (2016).

\bibitem{Marra2016}
P. Marra, R. Citro and A. Braggio, \prb\ {\bf 93}, 220507(R) (2016).

\bibitem{Li2016}
X. Li, B. Roy and S. Das Sarma, \prb\ {\bf 94}, 195144 (2016).

\bibitem{Baireuther2016b}
P. Baireuther, J. Tworzydlo, M. Breitkreiz, I. Adagideli and C. W. J. Beenakker, New J. Phys. {\bf 19}, 025006 (2017).


\bibitem{Madsen2016}
K. A. Madsen, E. J. Bergholtz and P. W. Brouwer, Phys. Rev. B {\bf 95}, 064511 (2017).


\bibitem{Obrien2017}
O'Brien, T. E. and Beenakker, C. W. J. and Adagideli, Phys. Rev. Lett. {\bf 118}, 207701 (2017).

\bibitem{Khanna2017}
U.~Khanna, S.~Rao, and A.~Kundu,
Phys. Rev. B {\bf 95}, 201115 (2017).

\bibitem{Bovenzi2017}
N. Bovenzi, M. Breitkreiz, P. Baireuther, T. E. O'Brien, J. Tworzydlo, I. Adagideli, and C. W. J. Beenakker,
Phys. Rev. B {\bf 96}, 035437.

\bibitem{Mukherjee2017} D. K. Mukherjee, S. Rao and A. Kundu,  \prb~{\bf 96}, 161408 (2017).

\bibitem{Hsieh2009} 
D. Hsieh, Y. Xia, L. Wray, D. Qian, A. Pal, J. H. Dil, F. Meier, J. Osterwalder,  G. Bihlmayer, C. L. Kane, Y. S. Hor, R. J. Cava and M. Z. Hasan, Science {\bf 323}, 919 (2009).


\bibitem{Chen2009} Y. L. Chen, J. H. Chu, J. G. Analytis, Z. K. Liu, K. Igarashi, H. H. Kuo, X. L. Qi, S. K. Mo, R. G. Moore, D. H. Lu, M. Hashimoto, T. Sasagawa, S. C. Zhang, I. R. Fisher, Z. Hussain and Z. X. Shen, Science {\bf 329}, 659 (2010).

\bibitem{Zhu2011} J. J. Zhu, D. X. Yao, S. C. Zhang and K. Chang, Phys. Rev. Lett. {\bf 106}, 097201 (2011). 

\bibitem{Wray2011} L. A. Wray, S. Y. Xu, Y. Xia, D. Hsieh, A. V. Fedorov, Y. S. Hor, R. J. Cava, A. Bansil, H. Lin and M. Z. Hasan, Nature Phys. {\bf 7}, 32 (2011).

\bibitem{Xu2012} 
S. Y.  Xu, M. Neupane, C. Liu, D.  Zhang, A.  Richardella, L. A.  Wray, N.  Alidoust, M.  Leandersson, T. Balasubramanian, J.  Sanchez-Barriga, O. Rader, G.  Landolt, B.  Slomski, J. H.  Dil, J.  Osterwalder, T. R. Chang, H. T.  Jeng, H.  Lin, A.  Bansil, N.  Samarth and  M. Z.  Hasan, Nature Phys. {\bf 8}, 616 (2012).

\bibitem{Hasan2010} M. Z. Hasan and C. L. Kane, Rev. Mod. Phys. {\bf 82}, 3045 (2010).

\bibitem{Qi2011} X. L. Qi and S. C. Zhang, Rev. Mod. Phys. {\bf 83}, 1057 (2011).


\bibitem{Wehling2014} T. O. Wehling, A. M. Black-Schaffer and A. V. Balatsky, Adv. Phys. {\bf 76}, 1 (2014).

\bibitem{Roy2016a}  S. Roy and S. Das, \prb{\bf 93}, 085422 (2016).

\bibitem{Roy2016b} S. Roy, K. Roychowdhury and S. Das, New J. Phys. {\bf 18}, 073038 (2016).

\bibitem{FanZhang}  F. Zhang, C. L. Kane and E. J. Mele, \prb ~{\bf 86}, 081303(R) (2012).



\bibitem{wu2010} Z. Wu, F. M. Peeters and K. Chang, \prb ~{\bf 82}, 115211 (2010).

\bibitem{campos2012} L. C.  Campos {\it et al},   Nat. Commun. 3, 2243 (2012).

\bibitem{oksanen2014} M. Oksanen {\it  et al},   Phys. Rev. B {\bf 89}, 121414(R) (2014).

\bibitem{zhang2013}
F. Zhang, C. L. Kane and E. J. Mele,  Phys. Rev. Lett. {\bf 110}, 046404 (2013).

\bibitem{Franz2016} D.I. Pikulin, Anffany Chen, M. Franz, Phys. Rev. X {\bf 6}, 041021 (2016)

\bibitem{bansal2011} N. Bansal {\it et al}, Epitaxial growth of topological insulator Bi2Se3 film on Si(111) with atomically sharp interface, Thin Solid Films,   {\bf 520}, 224 (2011).

\bibitem{kim2017}
{S. H. Kim {\it et al},
Atomically Abrupt Topological p-n Junction,  ACS Nano {\bf 11}, 9671 (2017).}


\end{thebibliography}
\end{document}